\documentclass{elsart}
\usepackage{epsfig}
\usepackage{amssymb}
\begin{document}
\begin{frontmatter}
\title{Can dissipation prevent explosive decomposition in high-energy
heavy~ion~collisions?}
\author[label1]{E. S.~Fraga}
\author[label2]{and G. Krein}
\address[label1]{Instituto de F\'\i sica, Universidade Federal
do Rio de Janeiro\\
Caixa Postal 68528, 21941-972 Rio de Janeiro, RJ  Brazil}
\address[label2]{Instituto de F\'{\i}sica Te\'orica, Universidade
Estadual Paulista \\
Rua Pamplona, 145, 01405-900 S\~ao Paulo, SP - Brazil}

\begin{abstract}
We discuss the role of dissipation in the explosive spinodal
decomposition scenario of hadron production during the chiral
transition after a high-energy heavy ion collision. We use a
Langevin description inspired by microscopic nonequilibrium field
theory results to perform real-time lattice simulations of the
behavior of the chiral fields. We show that the effect of
dissipation can be dramatic. Analytic results for the short-time
dynamics are also presented.
\end{abstract}


\end{frontmatter}

Lattice QCD results \cite{Laermann:2003cv} suggest that strongly
interacting matter at sufficiently high temperature undergoes a
phase transition (or a crossover) to a deconfined quark-gluon
plasma (QGP). Despite the difficulties in identifying clear
signatures of a phase transition in ultrarelativistic heavy ion
collisions, recent data from the experiments at BNL-RHIC clearly
point to the observation of a new state of matter \cite{QM2004}.

Depending on the nature of the QCD phase transition, the process
of phase conversion in the expanding QGP generated in a
high-energy heavy-ion collision may proceed in a number of
different ways. In general, there is a competition between the
mechanisms of nucleation and spinodal decomposition \cite{review}.
Results from CERN-SPS and BNL-RHIC feature what has been called
\textit{sudden hadronization} \cite{sudden} or \textit{explosive
behavior} \cite{explosive} in the hadronization process of the
expanding QGP and seem to favor a fast (explosive) spinodal
decomposition. Recently, possible signatures of this behavior in
high-energy nuclear collisions were proposed~\cite{randrup}.

Most theoretical attempts to understand this behavior focused on
the rapid changes in the effective potential of QCD near the
critical temperature such as predicted, for instance, by the
Polyakov loop model \cite{polyakov}, which would be followed by a
very fast spinodal decomposition process \cite{explosive}.  In
this case, one observes a nearly instantaneous decay of the
Polyakov loop condensate, producing an exploding source of pions
as well as large RMS spatial fluctuations in the chiral fields.
Previous studies using low-energy chiral effective models to
investigate how nucleation rates compare to the time scale of
expansion of the plasma also found very likely that most of the
system would reach the spinodal region and then undergo an
explosive phase
conversion~\cite{Csernai,Zab,Sca99,Sca01,Shukla,Paech}.

Therefore, both experiment and theory seem to suggest that a QGP
generated in a high-energy heavy ion collision expands and cools
down so fast that the processes of chiral symmetry breakdown and
hadronization can be described within the framework of an
effective potential that is quenched into the spinodal region,
where long-wavelength fluctuations grow with no barrier to
overcome. This leads to what we will refer to as the explosive
spinodal decomposition scenario. However, we will argue that even
if the system quickly reaches this unstable region there is still
no guarantee that it will explode. To asses the different
possibilities one has to study the time evolution of the order
parameter of the transition after the system is quenched beyond
the mean-field analysis, in order to incorporate the effects from
dissipation, and ask how they could modify the explosive
decomposition picture. In particular, dissipation effects have
proved to be important in the context of disoriented chiral
condensate (DCC) formation in heavy ion
collisions~\cite{biro,dirk,krishna,agnes,dcc}.

In this Letter we present an exploratory investigation of the
effects of dissipation in the explosive spinodal decomposition
scenario of hadron production during the QCD transition after a
high-energy heavy ion collision. We use a Langevin description
inspired by microscopic nonequilibrium field theory results to
perform real-time lattice simulations for the behavior of the
inhomogeneous chiral fields. We show that the effects of
dissipation can be dramatic even for very conservative
assumptions. Analytic results for the short-time dynamics are also
presented and discussed.

For the sake of simplicity, we consider an infinite system that is
quenched to the spinodal and then evolves with a fixed effective
potential. We believe that the subsequent evolution of the
effective potential should not bring deep modifications to our
general conclusions once the system has already reached the
unstable spinodal region. On the other hand, effects from the
finite size of the plasma will most likely play a non-trivial role
\cite{Fraga:2003mu}. Both effects will be taken into account in a
future publication \cite{future}.

In what follows, we consider the real-time dynamics of chiral
symmetry breakdown of a QGP created in a high-energy heavy ion
collision. We assume that the system is characterized by a
coarse-grained free energy
\begin{equation}
F(\phi,T)=\int d^3x \left[ \frac{1}{2}(\nabla\phi)^2 +
U_{eff}(\phi,T)  \right] \; , \label{free-energy}
\end{equation}
where $U_{eff}(\phi,T)$ is an effective potential of the
Landau-Ginzburg form whose coefficients depend on the temperature,
and $\phi(\vec{x},t)$ is a real scalar field which plays the role
of an order parameter that is {\em not} conserved, such as the
chiral condensate. To model the mechanism of chiral symmetry
breaking found in QCD, we adopt the linear $\sigma$-model coupled
to quarks, whose standard Lagrangian can be found, for instance,
in Ref.~\cite{Sca01}. This approach is widely used in the
literature and its specificities imply no major limitations to our
main results.

\begin{figure}[hbt]
\begin{center}
\includegraphics[height=.4\textheight]{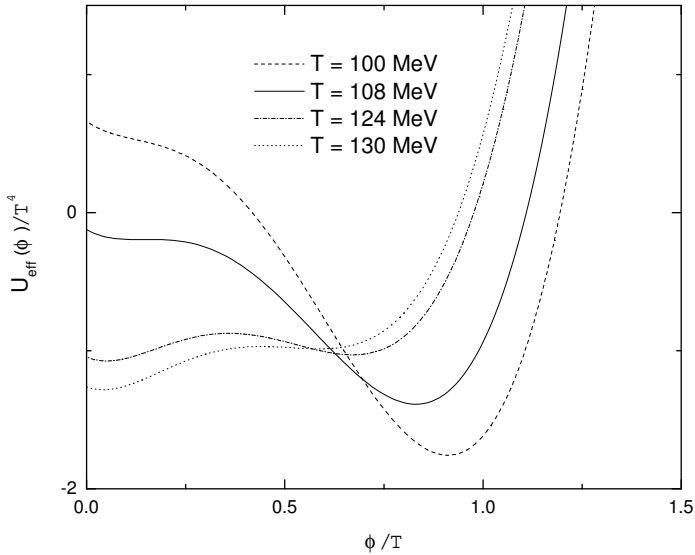}
\end{center}
\caption{Fit of the effective potential in the sigma direction for
different temperatures in the region of interest for the phase
conversion, $T_{sp} < T < T_c$.} \label{eff-pot-plot}
\end{figure}

Quarks can be treated as fast-moving modes and integrated out
using a classical approximation for the chiral field, yielding the
effective potential $U_{eff}(\phi,T)$ that is shown in
Fig.~\ref{eff-pot-plot}. The pion directions play no major role in
the process of phase conversion we are considering, so we
concentrate on the sigma direction represented by the field $\phi$
(see Ref.~\cite{Sca01} for details). In Fig. 1, we show
$U_{eff}(\phi,T)$ for a few values of temperature. The critical
temperature is $T_c\approx 123~$MeV and the spinodal is reached at
$T_{sp}\approx 108~$MeV. In order to simplify the numerics and to
compare with analytic results for the short-time behavior of the
order parameter evolution, we work with a fit of $U_{eff}(\phi)$
by a polynomial of sixth degree whose coefficients depend on the
temperature. Such a polynomial fit is an almost perfect
representation of $U_{eff}$, and it is the fit that is displayed
in Fig. 1.

In our analysis, the evolution of the order parameter
$\phi(\vec{x},t)$ and its approach to equilibrium will be dictated
by a Langevin equation of the form
\begin{equation}
\Box\,\phi + \Gamma \, \frac{\partial\phi}{\partial t} +
U'_{eff}(\phi)=\xi(\vec{x},t) \; , \label{langevin}
\end{equation}
where $\Gamma$, which can be seen as a response coefficient that
defines a time scale for the system and encodes the intensity of
dissipation, will be taken to be a function of temperature only,
$\Gamma=\Gamma(T)$. The function $\xi(\vec{x},t)$ represents a
stochastic (noise) force, assumed Gaussian and white, so that
$\langle \xi(\vec{x},t) \rangle = 0$ and $\langle \xi(\vec{x},t)
\xi(\vec{x'},t') \rangle = 2\,\Gamma T \delta(\vec{x}-\vec{x'})
\delta(t-t')$.

Eq. (\ref{langevin}) could, in principle, be obtained from a
microscopic field-theoretic description of the real-time
nonequilibrium dynamics of the chiral field at finite temperature.
This procedure was implemented in the case of a
$\lambda\phi^4$ scalar field theory in
Refs. \cite{Gleiser:1993ea,dirk}.
The noise and dissipation terms, which originate from quantum
fluctuations, are engendered by either self-interactions of the
chiral field or coupling to one or more different fields that play
the role of a heat bath, provided one incorporates higher order
terms in the computation of the effective equation of motion for
$\phi(\vec{x},t)$. In fact, it is well-known that one has to go up
to two-loop corrections in order to pick up imaginary parts in the
self-energy associated with viscosity and dissipation.
Self-interactions of the chiral field, as well as its interactions
with quarks and anti-quarks, justify the inclusion of a
dissipation and a noise term such as done in Eq. (\ref{langevin})
in the framework adopted in this paper. The same is true for an
approach that also includes the dynamics of the Polyakov loop
condensate coupled to the chiral field and, as we will argue, the
inclusion of this effect could dramatically modify the results of
Refs.~\cite{explosive,polyakov}.

The main physics behind $\Gamma$ is the decay of the $\phi$ field
and as such one would think that $\Gamma$ would be significant
only in regions where $\phi$ oscillates rapidly. However, this
needs not be the case in general. The procedure detailed in
Refs.~\cite{Gleiser:1993ea,dirk} leads to an effective equation of
motion that is richer in structure and much more complicated than
(\ref{langevin}). There is no {\it a~priori} reason for the
dissipation function $\Gamma$ be so simple and the noise be
Gaussian and white. In general, one obtains a complicated
dissipation kernel that simplifies to a multiplicative dissipation
term which depends quadratically on the amplitude of the field as
$\eta_1(T) \phi^2(\vec{x},t)\dot\phi(\vec{x},t)$, where $\eta_1$
is determined by imaginary terms of the effective action for
$\phi$ and depends weakly (logaritmically) on the coupling(s). The
fluctuation-dissipation theorem implies, then, that the noise term
will also contain a multiplicative contribution of the form
$\phi(\vec{x},t) \xi_1(\vec{x},t)$, and be in general
non-Markovian. The white-noise limit is reobtained only in the
limit of very high temperature. Nevertheless, assuming $\Gamma$ to
be a linear function of the temperature is a reasonable first
approximation, as can be seen from the results presented
in~\cite{dirk,krishna}. We will comment further on this point
later on when discussing our results.

For the sake of simplicity, we adopt the simple approximate form
of Eq. (\ref{langevin}) for a phenomenological description of the
dissipative evolution of the expectation value of the sigma field.
Although assumedly simple, this analysis allows for a clear
distinction and comparison of the roles played by dissipation and
the (explosive) spinodal instability in the spinodal decomposition
scenario of hadron production during the QCD transition after a
high-energy heavy ion collision. The simple form of Eq.
(\ref{langevin}) is also convenient for a comparison of our
numerical results to (linear) analytic estimates in the region of
short-time evolution to measure the effect of nonlinearities.

In our numerical simulations we solve Eq.~(\ref{langevin}) on~a
cubic space-like lattice with $64^3$ sites under periodic boundary
conditions, with a lattice spacing of $a~=~0.91$~fm. We use a
semi-implicit finite-difference scheme for the time evolution and
a Fast Fourier Transform for the spatial dependence~\cite{mimc}.
Temperature is fixed to the spinodal value $T_{sp}\approx
108~$MeV.  We perform several runs starting from different random
initial configurations around the inflexion point of $U_{eff}$
which happens at $\phi_0 \approx 0.162\,T$ and then average the
results from the different initial configurations. For time steps
of $\Delta t~=~0.001/T$ the results become independent of the
lattice spacing once it is smaller than $a~\simeq~1$~fm.

Before presenting the results of the simulations, it is
instructive first to analyze the short-time behavior of the
solution. One can linearize the equation around the inflexion
point $\phi_0$ substituting $\phi$ by $\phi = \phi' + \phi_0$ in
Eq.~(\ref{langevin}) and average over the noise. For short times,
$\phi'$ is small and the cubic and higher-order terms in
$U_{eff}(\phi')$ can be neglected, so that the equation for $\phi'$
becomes linear. Since the equation is linear, the average over the
noise can be done formally. An analytic form for the short-time
solution of the linear equation can be found by using the
polynomial fit of~$U_{eff}$,
\begin{equation}
U_{eff} = \sum_{n=0}^6 a_n\,\phi^n \;, \end{equation}
which leads to following equation for the average
$\langle\phi'\rangle$
\begin{equation}
\Box\,\langle\phi'\rangle + \Gamma
\frac{\partial\langle\phi'\rangle}{\partial t} + A \,
\langle\phi'\rangle = 0 \; , \label{lin-langevin}
\end{equation}
where $A=2 a_2 + 6 a_3 \phi_0 + 12 a_4 \phi^2_0 + 20 a_5\phi^3_0 +
30 a_6 \phi^4_0$. Note that the constant term in $U'_{eff}$ does
not contribute to the equation Eq.~(\ref{lin-langevin}) since
the average over noise of a constant is zero. One can write the
solution of Eq.~(\ref{lin-langevin}) in terms of the Fourier
transform $\langle{\widetilde\phi'}(\mathbf{k},t)\rangle$ of
$\langle\phi'(\mathbf{x},t)\rangle$ as
\begin{equation}
{\langle{\widetilde\phi}'(\mathbf{k},t\approx 0) \rangle} = C_1
e^{\lambda_1(\mathbf{k})\, t} + C_2 e^{- \lambda_2(\mathbf{k})\, t
}\;,
\end{equation}
where $C_1$ and $C_2$ are integration constants and
$\lambda_1(\mathbf{k})$ and $\lambda_2(\mathbf{k})$ are the roots
of the quadratic equation
\begin{equation}
\lambda^2(\mathbf{k}) + \Gamma \lambda (\mathbf{k}) + \left(
\mathbf{k}^2 - |A|\right)=0 \; ,
\end{equation}
where we used the fact that $A <0$ in our case. From this one sees
that for short wavelengths, such that $\mathbf{k}^2 \gg \Gamma^2/4
+ |A|$, we have (complex conjugate) imaginary roots and the
solution is oscillatory. For long wavelengths, such that
$\mathbf{k}^2 < \Gamma^2/4 + |A|$, we have real roots and there is
an exponential growth of the Fourier components. This exponential
growth yields the explosive spinodal decomposition.

Two other limits are also instructive. One is the strong
dissipation limit of large $\Gamma$, such that the first order
time derivative dominates over the second order one. In this case,
the short-time solution to Eq.~(\ref{lin-langevin}) is given by
\begin{equation}
{\langle{\widetilde\phi}'(\mathbf{k},t\approx 0) \rangle} = C_0 \,
e^{ - (\mathbf{k}^2 - |A|)\, t/\Gamma} \; . \end{equation}
In this case, one sees that short wavelengths with $\mathbf{k}^2
> |A|$ are absorbed by the system, while those with $\mathbf{k}^2
< |A|$ explode exponentially. Of course, as time increases,
$\phi'$ increases and the linear equation is not valid anymore and
the fully nonlinear equation has to be solved. The other
interesting limit is the one with $\Gamma =0$, i.e. no
dissipation. We will discuss this limit in connection with the
solution of the full nonlinear equation in the following.

We show results of simulations for three different values of the
dissipation coefficient, namely $\Gamma = 0$, $2\,T$ and $4\,T$.
It can be argued that the response coefficient has the form
$\Gamma(T) \approx 2\,T/b$, where $b$ is a number of order one to
first approximation \cite{Kajantie:1992uk}. The cases considered
provide a conservative band around the value $\Gamma(T) \approx
2\,T$ to illustrate the effect of dissipation.

\begin{figure}[hbt]
\begin{center}
\includegraphics[height=.4\textheight]{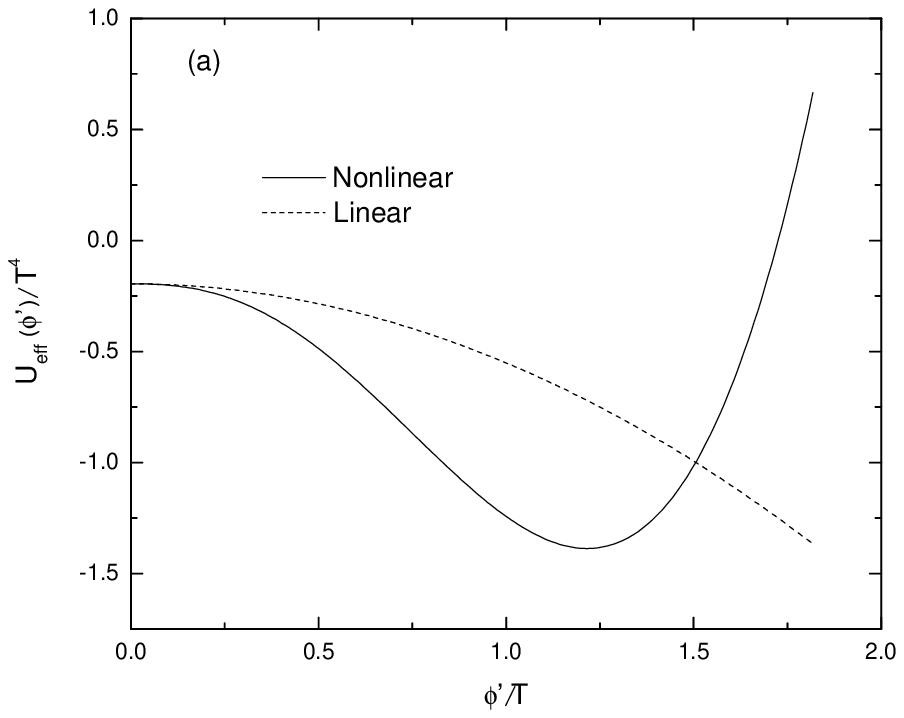}\vspace{-0.5cm}
\includegraphics[height=.4\textheight]{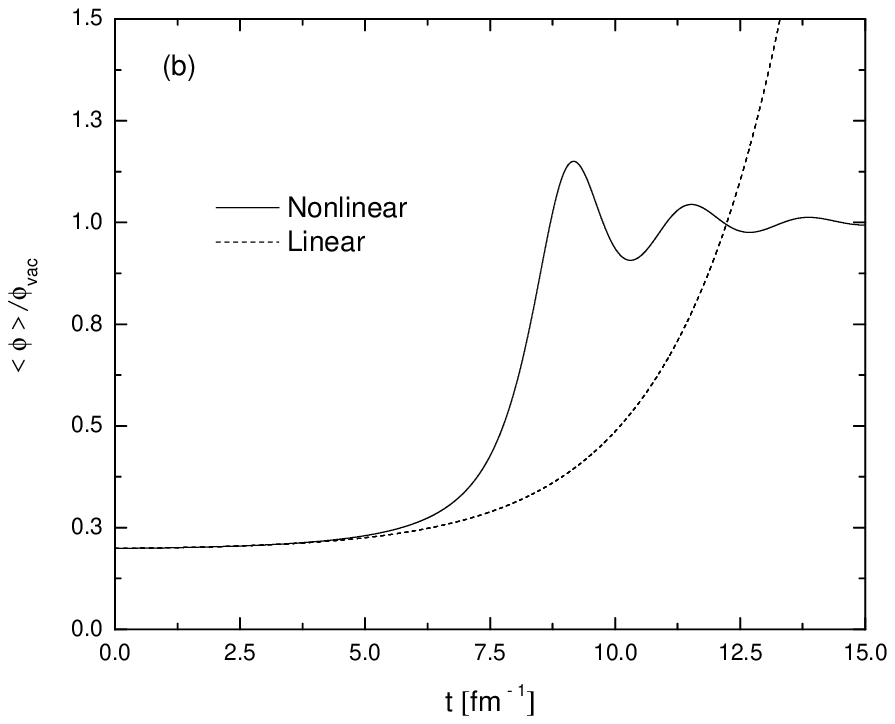}
\end{center}
\caption{(a) The full nonlinear (solid line) and the ${\mathcal
O}(\phi'^2)$ effective potential (dashed~line). (b) The
corresponding average values of $\phi$ in units of its vacuum
value $\phi_{vac}$ as a function of time for $\Gamma/T= 2$.}
\label{lin-nonl}
\end{figure}

In Fig.~\ref{lin-nonl} we compare the solutions of the full,
nonlinear equation with the solution of the linearized equation
for $\Gamma/T = 2$. Fig.~\ref{lin-nonl}~(a) shows that the
roll-down for the ${\mathcal O}(\phi'^2)$ potential is slower than
for the full $U_{eff}$ potential. This is obviously due to the
fact that the falloff of full $U_{eff}$ is steeper that that of
the ${\mathcal O}(\phi'^2)$ for small $\phi'$. At short times,
smaller than $t=5$~fm$^{-1}$,  one sees that both solutions are
very close to each other. One interesting aspect of the solutions
shown in Fig.~\ref{lin-nonl}~(b) is that the exponential explosion
of the linear solution happens much later than the explosion of
the full nonlinear equation. This seems at first sight very
counterintuitive since, from the discussion above, one would
expect an early-time explosion of the solution of linearized
solution. This does not happen here because the ${\mathcal
O}(\phi'^2)$ potential is much shallower than the full $U_{eff}$
for small $\phi'$.

\begin{figure}[htb]
\begin{center}
\includegraphics[height=.4\textheight]{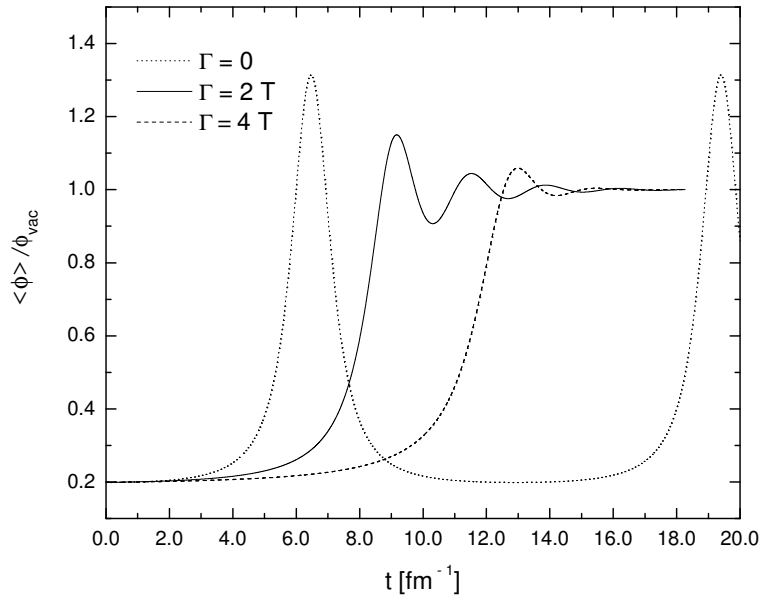}
\end{center}
\caption{Average value of  the chiral field $\phi$ in units of its
vacuum value $\phi_{vac}$ as a function of time for $\Gamma/T= 0,
2, 4$. } \label{phi-ave}
\end{figure}

In Figure~\ref{phi-ave} we show the average value of $\phi$ in
units of its vacuum value, $\phi_{vac}$, as a function of time for
the three different values of $\Gamma$ mentioned above. In the absence
of dissipation, the solution (dotted line) is obviously
oscillatory, with an explosive early-time behavior. The effect of
dissipation is of retarding the exponential growth, as shown by
the solid and dashed lines in the Figure.

The results clearly show that even for a very conservative value
of dissipation, $\Gamma = 2\,T$, the effect can be dramatic. For
this value of $\Gamma$, dissipation retards the time evolution of
$\phi$ towards its vacuum value in $\sim 100\%$ compared to the
case with $\Gamma=0$. The important point to be noted here is that
for expansion times of the order of $5~$fm$^{-1}$, which in of the
order of the time scales for the RHIC collisions, there might be
not enough time for the onset of the spinodal
explosion~\cite{explosive}.

Of course, effects brought about by the expansion of the
plasma \cite{explosive}
and by its finite size \cite{Fraga:2003mu}, as well as a more
realistic treatment of dissipation from the microscopic point of
view, will bring corrections to this picture.
For instance, the authors of Ref. \cite{explosive} consider a
Hubble expansion of the system which introduces a
dissipation-like term to the evolution equation of the
form $H (\partial\phi / \partial\tau$), where $\tau$ is the proper
time and $H = 1/\tau$ is the expansion rate. Therefore, for a very
rapid expansion, corrections due to dissipation, such as discussed here,
should play a comparatively less important role.
Also, dissipation being mainly the result of the decay of the $\phi$
field, common wisdom would suggest that its effect should be less
important at short times, when the field is slowly starting to
roll down the potential. However, as discussed earlier, $\Gamma$
depends, in general, on $\phi$ - in a $\lambda \phi^4$ model it
is proportional to $\phi^2$ - and it is not a priori clear
what will be the effect of such a dependence on the short-time
evolution of the system.
These issues will be addressed in a future publication
\cite{future}.

\vspace{1.0cm}
\noindent {\bf Acknowledgments}

\noindent E.S.F. would like to thank F. Gelis and E. Iancu for
their kind hospitality at the Service de Physique Th\'eorique,
CEA/Saclay, where part of this work has been done.
The authors are grateful to A. Dumitru for a critical reading of
the manuscript and several suggestions.
The work of E.S.F. is partially
supported by CAPES, CNPq, FAPERJ and FUJB/UFRJ. The work of G.K.
is partially supported by CNPq and FAPESP.

\end{document}